\documentclass[prl,twocolumn,showpacs,superscriptaddress,nofootinbib]{revtex4-2}
\pdfoutput=1
\usepackage{graphicx}
\usepackage{epstopdf}
\usepackage{bm}
\usepackage{amssymb}
\usepackage{color}
\usepackage{amsmath}
\usepackage{amstext}
\usepackage{latexsym}
\usepackage[usenames,dvipsnames]{xcolor}
\usepackage[colorlinks=true,citecolor=MidnightBlue,linkcolor=RubineRed,urlcolor=MidnightBlue]{hyperref}

\def\x{{\rm\bf x}}

\newcommand{\beq}{\begin{equation}}
\newcommand{\eeq}{\end{equation}}
\newcommand{\beqa}{\begin{eqnarray}}
\newcommand{\eeqa}{\end{eqnarray}}

\usepackage{tikz,xcolor,hyperref}
\definecolor{lime}{HTML}{A6CE39}
\DeclareRobustCommand{\orcidicon}{
\begin{tikzpicture}
\draw[lime, fill=lime] (0,0)
circle[radius=0.16]
node[white]{{\fontfamily{qag}\selectfont \tiny \.{I}D}}; 
\end{tikzpicture}
\hspace{-2mm}
}
\foreach \x in {A, ..., Z}{%
\expandafter\xdef\csname orcid\x\endcsname{\noexpand\href{https://orcid.org/\csname orcidauthor\x\endcsname}{\noexpand\orcidicon}}
}

\begin{document}

\title{Emergence of Large-Scale Structures in Holographic Superfluid Turbulence}

\author{Wei-Can Yang\hspace{-1.5mm}\orcidA{}
}
\affiliation{Department of Physics, Osaka Metropolitan University, 3-3-138 Sugimoto, 558-8585 Osaka, Japan}

\author{Chuan-Yin Xia}
\affiliation{Center for Theoretical Physics , Hainan University, Haikou 570228, China}
\affiliation{Center for Gravitation and Cosmology, College of Physical Science and Technology, Yangzhou University, Yangzhou 225009, China}

\author{Yu Tian}\email{ytian@ucas.ac.cn}
\affiliation{School of Physical Sciences, University of Chinese Academy of Sciences, Beijing 100049, China $\&$ Institute of Theoretical Physics, Chinese Academy of Sciences, Beijing 100190, China}

\author{Makoto Tsubota}\email{tsubota@omu.ac.jp}
\affiliation{Department of Physics, Osaka Metropolitan University, 3-3-138 Sugimoto, 558-8585 Osaka, Japan}
\affiliation{Nambu Yoichiro Institute of Theoretical and Experimental Physics (NITEP), Osaka Metropolitan University, 3-3-138 Sugimoto, Sumiyoshi-ku, Osaka 558-8585, Japan}

\author{Hua-Bi Zeng\hspace{-1.5mm}\orcidB{}
}\email{zenghuabi@hainanu.edu.cn}
\affiliation{Center for Theoretical Physics , Hainan University, Haikou 570228, China}
\affiliation{Center for Gravitation and Cosmology, College of Physical Science and Technology, Yangzhou University, Yangzhou 225009, China}


\begin{abstract}
In two-dimensional turbulence systems, 
the emergence of large-scale structures holds profound physical implications, particularly as it indicates the occurrence of inverse energy cascades, thereby garnering significant attention.
In this paper, we report a novel vortex clusters formation in the background of near-extreme Reissner-Nordstr$\ddot{o}$m black hole holographic model. At temperatures nearing absolute zero, we observe not only the formation of vortex clusters but also the emergence of an inverse energy cascade. Distinct from typical quantum systems, the genesis of holographic vortex clusters is rooted in unique quantum dissipation properties, characterized by the near immobilization of vortex dipoles at low temperatures. Through a comparative analysis with the dynamics of the Gross-Pitaevskii equation, our investigation enhances the understanding of inverse energy cascades under these extreme conditions, thereby broadening our comprehension of quantum turbulence.
\end{abstract}

\maketitle
{\it Introduction.}
The intricate and intense nonlinear dynamics of classical turbulence present a challenge in developing a unified model to describe its chaotic behavior. This unpredictability is largely caused by the chaotic movements of vortices across various scales, which are central to the turbulence phenomenon \cite{Saffman1981,Frisch1995,Davidson2015}.
Uncovering specific patterns within the unpredictable has always been a key endeavor for physicists, and a prime example of this is Lars Onsager's 1949 prediction of the negative temperature states for vortex system \cite{Onsager_1949}. 
The Onsager model suggests that in a two-dimensional point vortex system, when viscous dissipation is negligible, the system tends to maintain high-energy equilibrium, allowing the system to reach a state of inverse energy distribution. In this state, larger vortices in the system absorb more energy, causing them to move faster, while smaller vortices release energy. This method of energy distribution implies that the ``temperature" of the system can be defined as negative, because an increase in energy leads to a decrease in the system's entropy, which is the opposite of what happens in positive temperature systems where an increase in energy leads to an increase in entropy.  
However, due to the continuous distribution of vortices in classical turbulence, it is challenging to simulate and achieve a quantitative understanding in practice.
Conversely, quantum turbulence offers an ideal experimental platform as its quantized vortices provide a perfect approximation to the discrete model \cite{Feynman_1955,Barenghi_2023,Tsubota_2013}.
Numerous numerical and theoretical studies focusing on the relaxation dynamics of vortices have revealed the formation of Onsager vortices in non-dissipative two-dimensional quantum turbulence at zero temperature, utilizing the Gross-Pitaevskii equation \cite{Simula2014,Groszek2018,Valani_2018,Han2019,Groszek2016,Kanai2021,Reeves2013,Billam2015,Billam2014}. These studies have also confirmed the presence of an inverse energy cascade in the energy spectrum \cite{Reeves2013,Billam2015,Billam2014}.
More exhilaratingly, in recent cold-atom experiments, additional energy was imjected to zero-temperature vortex systems through vortex evaporation heating and transient stirring, leading to the formation of transient vortex cluster states and the sustainment of negative temperatures.\cite{Johnstone_2019,Gauthier_2019}.


The success observed in weakly interacting systems raises the anticipation of discovering large-scale structures within strongly interacting systems. 
However, the high complexity brought about by strong correlations has limited progress both theoretically and numerically. Fortunately, in recent years, the discovery of holographic duality in string theory, which maps quantum problems onto a higher-dimensional gravity problem, has made it possible to solve complex dynamic problems in strongly interacting systems \cite{Maldacena_1999,Gubser_1998,witten1998}. 
Since the establishment of the holographic superconductor/superfluid model \cite{Hartnoll2008,Herzog2009}, holographic models have achieved remarkable results in the field of condensed matter \cite{Chesler_2013,Guo2020,Wittmer2021,xia2021,Lan2023,Yang2023Universal,yang2024spontaneous}, especially near the phase transition critical points, where they align well with the outcomes predicted by Landau mean-field theory \cite{Yin2015}. 
Recently, studies employing holographic superfluids to investigate low-temperature vortex dynamics have unveiled important discoveries that extend beyond the weakly coupled regime \cite{Yang2023}. Consequently, although research on holographic superfluid turbulence at higher temperatures suggests that inverse energy cascades might not be a characteristic feature of holographic turbulence \cite{Chesler_2013,Lan2016}, it is imperative to further examine their potential in the regime of temperatures approaching absolute zero.

In this Letter, we investigate the dynamical evolution of holographic superfluid turbulence at temperatures approaching absolute zero, and present a comparative study with the turbulence evolution governed by the Gross-Pitaevskii equation. Our observations reveal that the vortex motion within these two frameworks exhibits distinctly different behaviors, yet intriguingly, both lead to the emergence of vortex clusters. This marks the first reported instance of vortex cluster formation and inverse energy cascades within the background of holographic models, and more generally, in the strongly coupled systems.

{\it Model.} 
We conducted simulations of superfluid vortex turbulence within a two-dimensional framework, incorporating a circular confining potential. 
In our approach to the holographic model, we utilize a (2+1)-dimensional superfluid system that is dual to a (3+1)-dimensional asymptotically anti-de Sitter (AdS) black hole spacetime. This application of holographic duality enables us to explore the dynamics of lower-dimensional, strongly correlated quantum field theory by solving the higher-dimensional, weakly correlated gravitational theory. 
The action in the higher-dimensional bulk is formulated as follows:
\begin{equation}
    S=\int d^4x \sqrt{-g} \big[-\frac{1}{4}F_{\alpha\beta} F^{\alpha\beta}-|\partial_\alpha \Psi-iqA_\alpha \Psi|^2-m^2|\Psi|^2 \big]
\end{equation}
where $F_{\alpha\beta}=\partial_\alpha A_\beta-\partial_\beta A_\alpha$ is the Maxwell field strength with vector potential $A_{\alpha}$, that is coupled  to the complex scalar field $\Psi$ involving the charge $q$ and the mass has value of $m^2=-2$.  

The metric tensor $g$ of the gravitational background is taken from the Reissner-Nordstr$\ddot{o}$m black hole \cite{Liu2011,Chamblin1999,Romans_1992}
\begin{align}
      ds^2=\frac{1}{z^2}[-f(z)dt^2-2dtdz+dx^2+dy^2]\\
   f(z)=1+Q^2(z)^4-(1+Q^2)(z)^3 
\end{align}
the Hawking temperature is given by $T=(3-Q^2)/4\pi$. In the special case of $Q=0$,  the gravitational background reverts to that of a standard Schwarzschild black hole, with its temperature governed by the non-dimensional chemical potential \cite{Hartnoll2008}.
Accordingly, we establish the probe chemical potential at the superfluid phase transition's critical point, setting $A_t|_{z=0}=\mu=\mu_c=4.07$ \cite{Yang2023}. 
Therefore, varying $Q$ within the range from $0$ to $\sqrt{3}$ allows us to model the system's transition from the critical temperature of the superfluid phase to absolute zero. In this study, we select $Q^2=2.985$, corresponding to a temperature of $T=0.005T_c$, which is exceedingly close to absolute zero, thereby allowing us to probe the dynamics of superfluid turbulence under near-zero temperature conditions.

\begin{figure}[t]
    \centering
    \includegraphics[width=10cm,trim=25 0 0 0]{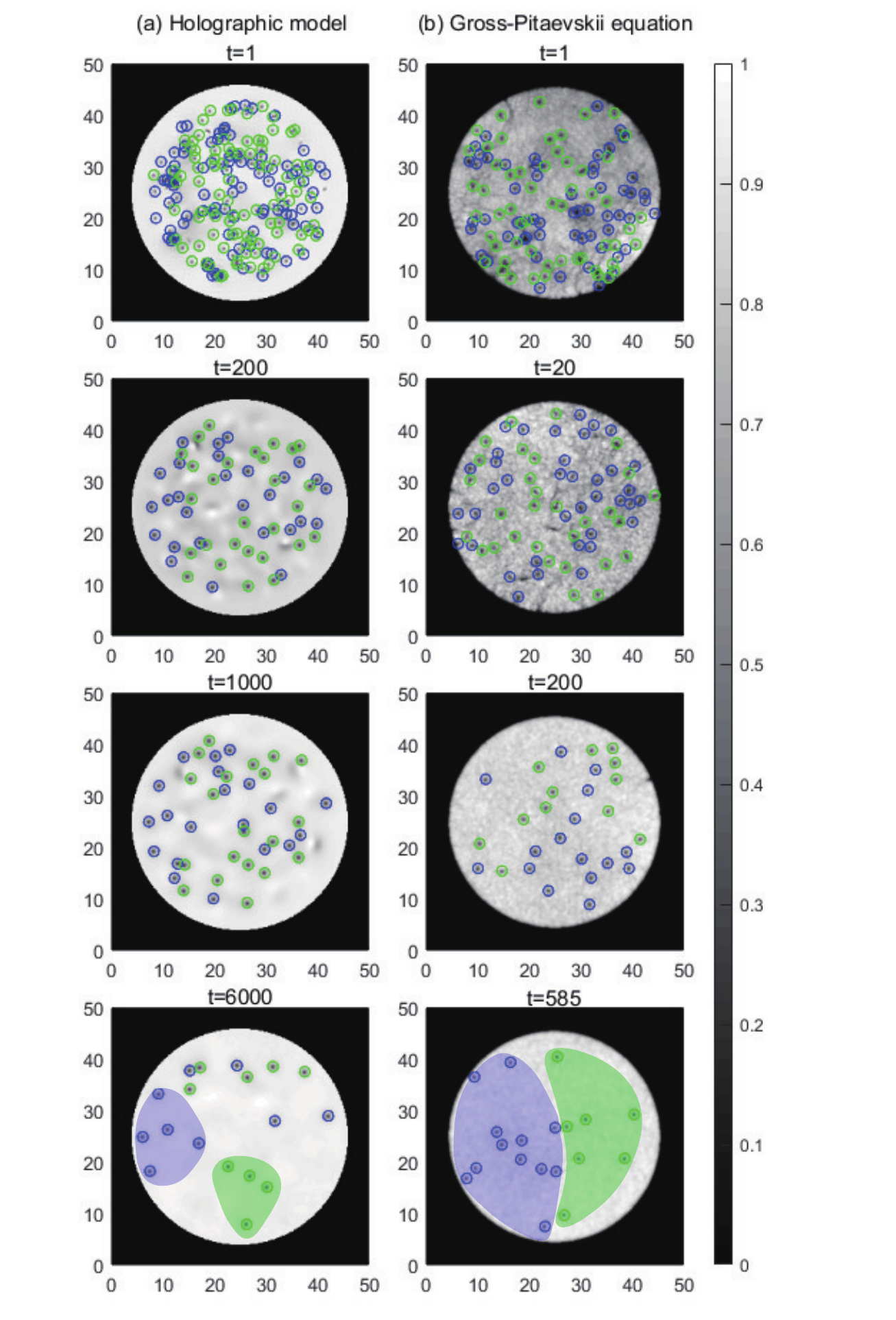}
    \caption{The superfluid density diagram simulated by (a) holographic model and (b) Gros-Pitaevskii equation for the evolution process of vortex relaxation dynamics. The density is normalized by the initial density $\rho/\rho_i$, and the snapshots at different times are represented from top to bottom. The positive vortices are represented by green circles, while the anti-vortices are represented by blue ones. In the last image, the vortex clusters are framed with color blocks. 
For more detailed and intuitive observations, please refer to the videos in the supplementary materials \cite{Supplementary}.}
    \label{figure1}
\end{figure}

In our comparative analysis, we utilize the zero-temperature two-dimensional Gross-Pitaevskii equation. The kinetic equation governing this system is given by:
\begin{equation}
        i\hbar\frac{\partial}{\partial t}\Psi = (-\frac{\hbar^2}{2m}\nabla^2+V_{trap}+g|\Psi|^2-\mu)\Psi
\end{equation}
where $m$ is the mass of an atom, $V_{trap}$ is the trapping potenetial which radially confines the condensate, $g$ is the interaction parameter and $\mu$ is the chemical potential.
To maintain generality, we adopt dimensionless units: $\hbar=m=g=1$. The healing length is defined as $\xi=\hbar/\sqrt{m\mu}$ with our setting $\mu=0.3$.

{\it Simulation and analysis.} 
In both models, we employ a circular confining potential with a radius of 18 to achieve non-periodic boundary conditions in a periodic superfluid box \cite{Lan2023}. To initiate the dynamics, we randomly imprint vortices by multiplying phase factor $\prod^{N_v}_{i=1}\mathrm{exp}(i\phi_i)=\prod^{N_v}_{i=1}\mathrm{exp}(is_i \mathrm{arctan}[(y-y_i)/(x-x_i)])$ on the global scalar field $\Psi(z)=|\Psi(z)| e^{i\phi}$ \cite{Han2019}. The coordinates $(x_i,y_i)$ refer to the position of the i-th vortex and $s_i= \pm 1$ corresponds to the winding number of the vortex.
We prepared 60 randomly distributed positive vortices and negative vortices, respectively.

Following a brief period of evolution, the vortices begin to emerge and initiate movement. This dynamic process is illustrated in Fig.~\ref{figure1}, where the left side displays the results from the holographic superfluid simulation, and the right side presents the outcomes derived from the Gross-Pitaevskii equation. There are corresponding movies in the supplementary material \cite{Supplementary}. 
Notably, the two models exhibit distinctly different dynamic behaviors. In the holographic model, the movement of vortices is gentle, with minimal changes in their relative positions. This behavior aligns with the motion characteristics of vortices in ultra-low temperature holographic superfluid \cite{Yang2023}, epitomizing the peculiar thermalization behavior of vortices in strange metallic states. Conversely, in the Gross-Pitaevskii equation simulation, vortices exhibit more vigorous movements, and in the absence of dissipation, they do not annihilate except upon direct collision.

\begin{figure}[t]
    \centering
    \includegraphics[width=9cm,trim=0 0 0 0]{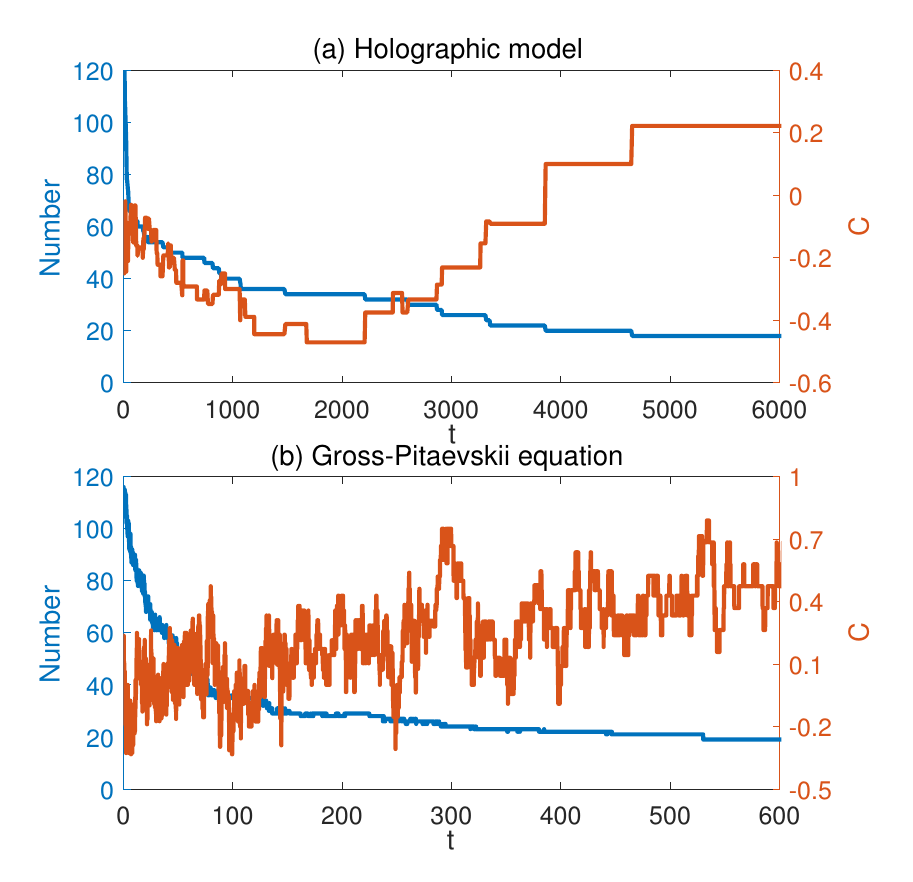}
    \caption{The time evolution of vortex number and vortex cluster correlation function in the simulation of (a) holographic model and (b) Gross-Pitaevskii equation, where the blue line represents the vortex number and the red line represents the correlation function.}
    \label{figure2}
\end{figure}

Interestingly, after a sufficient period of evolution, both models exhibit the formation of local clusters of vortices with the same circulation. However, the differing vortex dynamics between the two models suggest that the underlying mechanisms driving cluster formation are not identical.

In Fig.~\ref{figure2}, we show the time evolution of vortex number $N$ and vortex cluster correlation function $C$. The vortex cluster correlation function is defined as $C=\frac{1}{N}\sum^N_{i=1}c_i$, where $N$ is the total vortex number, and $c_i=\pm 1$ when the circulation of the nearest neighbour of the i-th vortex 
has the same sign or opposite sign \cite{Panico_2023}. Therefore, $C<0$ means that the vortex is basically in the form of a dipole, while $C>0$ means that the vortex clusters dominate.
Similar evolutionary trends are observed in both the holographic model and the Gross-Pitaevskii equation. Following an initial rapid phase of vortex annihilation, the vortex count tends to stabilize, predominantly leaving behind those vortices with higher energy levels that are more challenging to annihilate. The persistence of a positive correlation function, when the vortex count is stable, indicates the formation and enduring presence of vortex clusters.

\begin{figure}[t]
    \centering
    \includegraphics[width=9.5cm,trim=25 0 0 0]{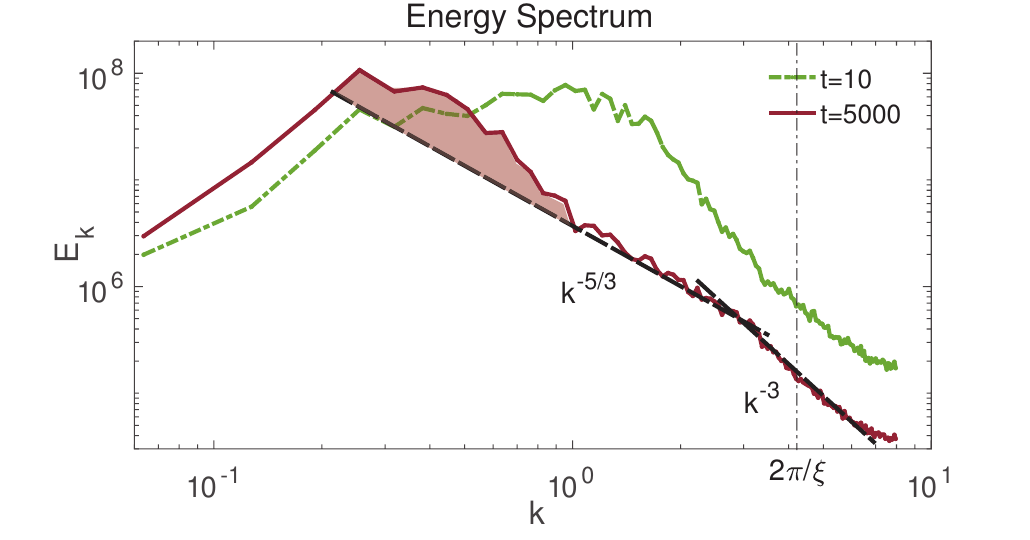}
    \caption{Energy spectrum during the formation of holographic vortex clusters. The green dotted line represents the spectrum when $t=0$ with no vortex clusters, the red solid line is the spectrum when $t=5000$ with vortex clusters stabilized, the two black dotted lines represent $-5/3$ law and $-3$ law, and the dotted line in the middle is the wavelength associated with the correlation length $\xi$. The red block are spectral condensations associated with large-scale structures.}
    \label{figure3}
\end{figure}

The emergence of large-scale structures is frequently associated with inverse energy cascades, a phenomenon observed in the formation of vortex clusters within the Gross-Pitaevskii framework as well \cite{Reeves2013}. To investigate this further in the holographic model, we examined the energy spectrum. In the holograhic model, the energy spectrum can be calculated from the wave function $\psi$:
\begin{eqnarray}\label{energyspectrum}
E_k(k)=\frac{1}{2}\int^\infty_0 d\theta k \mathcal{V}^{\star} ({\bf k}) \mathcal{V}({\bf k}),
\end{eqnarray}
where $\mathcal{V}=\langle \psi \rangle {\bf u}$ and ${\bf u}$ is the superfluid velocity $\frac{i}{2}[\langle \psi^{\star} \rangle \nabla \langle \psi \rangle - \langle \psi \rangle \nabla \langle \psi \rangle ^\star ] / |\langle \psi \rangle|^2$.
as depicted in Fig.~\ref{figure3}. We have compared the energy spectrum from an initial phase, prior to the formation of vortex clusters, with a subsequent phase when the vortex clusters has stabilized. It is evident that energy flows from smaller scales $(k>1/\xi)$ towards larger scales $(k<1/\xi)$, showcasing a distinct inverse energy cascade phenomenon.
Moreover, the energy spectrum adheres to the Kolmogorov $-5/3$ law, indicative of energy flux requirements, and displays the $-3$ law characteristic within individual vortices. These observations further corroborate the presence of turbulence dynamics, where the -5/3 law pertains to the inertial subrange of the energy spectrum in fully developed turbulence \cite{Kobayashi2005}, and the -3 law is often associated with the spectral distribution within discrete vortex structures \cite{Bradley2012}.
In the long-wavelength region, a hump appears in the energy spectrum, corresponding to spectral condensation induced by the presence of vortex clusters \cite{Simula2014}.
\begin{figure}[t]
    \centering
    \includegraphics[width=9.3cm,trim=70 30 0 10]{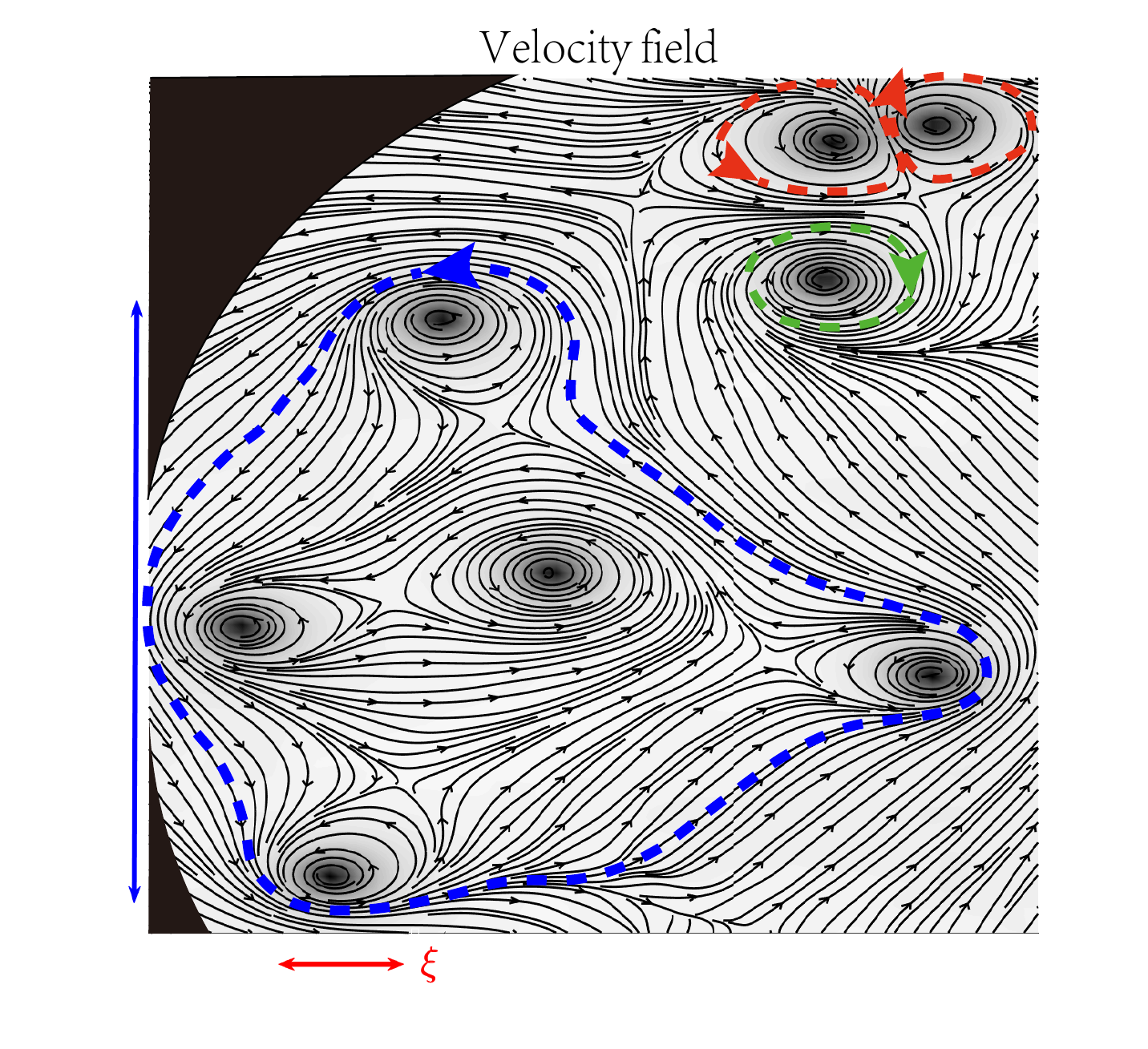}
    \caption{Local magnification of superfluid density, and distribution of velocity field at $t=6000$. The blue line frames the large-scale velocity field of the vortex cluster, and the red and green frame the velocity field of the dipole and the individual vortex, respectively. The blue double arrow line indicates the scale of the vortex cluster, the red double arrow line indicates the correlation length.}
    \label{figure4}
\end{figure}

In order to emphasize the macroscopic mechanism of energy transfer to large scales, in Fig.~\ref{figure4}, we enlarge the local structure of the superfluid turbulent flow and give the velocity distribution consisting of a vortex cluster of five vortices, a dipole, and a single vortex. It is evident that within the vortex cluster, the velocity field is organized on a larger scale, circulating around the five vortices. In contrast, for dipoles and individual vortices, the circulation occurs only on a smaller scale within correlation length $\xi$. 

\begin{figure}[t]
    \centering
    \includegraphics[width=8.5cm,trim=0 0 0 0]{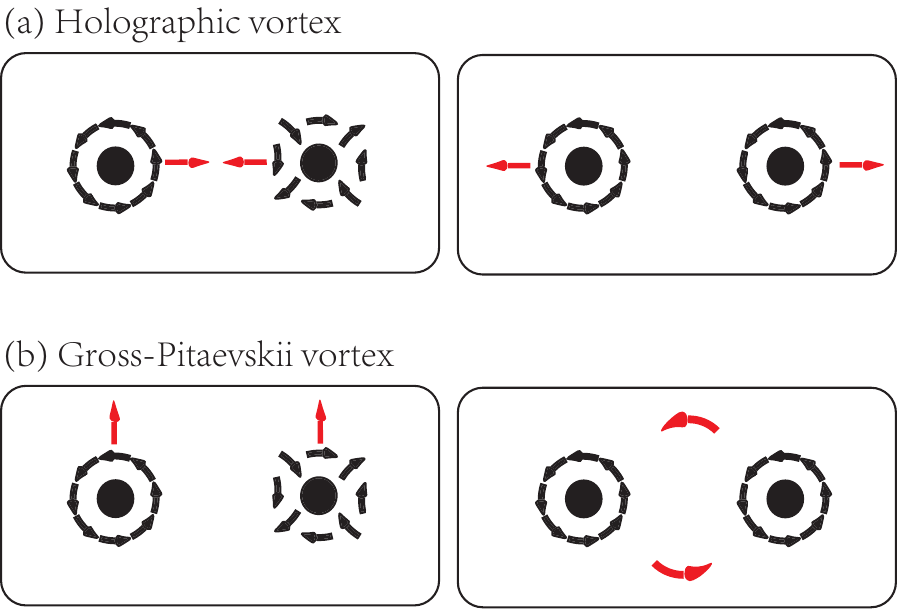}
    \caption{The motion laws of the (a) holographic vortex pair and (b) Gross-Pitaevskii vortex pair at zero temperature.}
    \label{figure5}
\end{figure}

Based on the comprehensive analysis conducted earlier, including the examination of superfluid density, correlation functions, energy spectrum, and velocity fields, we have observed the formation of vortex clusters and inverse energy cascades within holographic superfluid turbulence. 
However, we know that although vortex clusters are obtained in the holographic model, their formation process is completely different from that in the Gross-Pitaevskii equation. It is generally believed that due to the so-called evaporative heating mechanism in the Gross-Pitaevskii equation \cite{Simula2014}, that is, the collision of vortices annihilates the vortices with lower energy, leaving the vortices with higher energy, so that the system tends to separate vortices to maintain a higher energy state. However, in the holographic model, we do not observe the violent movement and collision annihilation of the vortices, and the vortices almost stay near the initial position.
By examining the dynamic evolution of vortex dipoles, we summarize the difference between the motion laws of holographic and Gross-Pitaevskii vortex pair in Figure.\ref{figure5}. At zero temperature, in the Gross-Pitaevskii equation, vortex dipoles move in parallel, with vortices of the same sign rotating around a central point, resulting in vigorous motion. 
However, in holographic superfluid, due to the influence of strange metal core \cite{Yang2023}, the vortices with different sign move slowly relative to each other, and the vortices with the same sign move in opposite directions, resulting in the slow evolution of the whole. And because only the relative motion of the dipoles with the opposite sign eventually annihilates, while the vortices with the same sign become stationary due to the decrease of interaction after the opposite motion reaches a certain distance, the vortex cluster with the same sign eventually forms.

{\it Conclusion.}
We report the first observation of spontaneous formation of large-scale structures in holographic superfluid. Through detailed analyses of superfluid density, correlation functions, energy spectrum, and velocity fields, we have identified the formation of vortex clusters associated with inverse energy cascades. By comparing with simulations using the Gross-Pitaevskii equation and analyzing the dynamics of vortex dipoles, we provide an explanation for the emergence of this phenomenon in holographic superfluid turbulence. We anticipate that this discovery will deepen our understanding of superfluid turbulence and look forward to potential validation in future theoretical and experimental studies of strongly correlated systems.

{\it Acknowledgements.}
H.B. Z. acknowledges the support by the National Natural Science Foundation of China (under Grant No. 12275233). M. T. acknowledges the support by the
JSPS KAKENHI (under Grant No. JPK23K03305 and JPH22H05139). Y. T. acknowledges the support by the National Natural Science Foundation of China (under Grant No. 12035016, 12361141825 and 12375058).

\onecolumngrid
\appendix 
\clearpage
\section*{Supplementary Material}

In this supplementary material, we provide a detailed expansion on the description of the model and the application of numerical methods. 
Furthermore, we present simulation of the relaxation dynamics of vorteices at finite-temperature, for comparison with the near-zero temperature results discussed in the main text.

\section{1. Holographic superfluid model}

The holographic superfluid model is an innovative theoretical framework that utilizes the principles of holographic duality to study the properties and dynamics of superfluid systems. This model is grounded in the correspondence between a higher-dimensional gravitational theory and a lower-dimensional quantum field theory, offering a powerful tool for exploring the complex behaviors of superfluids, especially in strongly coupled regimes where traditional methods struggle. By embedding complex scalar fields and gauge fields into the Einstein-Maxwell action, the model captures the property of strongly-coupled superfluid, allowing for the examination of phenomena such as phase transitions, vortex dynamics, and non-equilibrium processes at both zero and finite temperatures. 

In our simulations of the relaxation dynamics involving a large number of vortices, we employed a bottom-up holographic model. The action for this model is formulated as a coupling between the Einstein-Hilbert action, which describes the gravitational sector, and the action for massive gauge fields
\begin{align}
S=\frac{1}{16\pi G_N} \int d^4x \sqrt{-g}\Big[\mathcal{L}_{EH}-\mathcal{L}_{matter}
\Big]
\\
\mathcal{L}_{EH}=R+6/L^2
\\
\mathcal{L}_{matter}=-\frac{1}{4}F_{\alpha\beta} F^{\alpha\beta}-|\partial_\alpha \Psi-iqA_\alpha \Psi|^2-m^2|\Psi|^2
\end{align}
where $\mathcal{L}_{EH}=R+6/L^2$ is the Einstein-Hilbert action with cosmological constant $6/L^2$ wiring in the asymptotic AdS geometry. $F_{\alpha\beta}=\partial_\alpha A_{\beta}-\partial_\beta A_{\alpha}$ is the Maxwell field strength with vector potential $A_{\alpha}$.  $\Psi$ is the complex scalar field with mass  $m$.

In this context, we can adopt the convention of setting $m^2=-2$ and $q=1$ without sacrificing the generality of our analysis. As per the holographic duality, the behavior of the bulk field near the boundary is characterized by $\Psi = \phi z + \psi z^2 + \mathcal{O}(z^3)$, while the gauge field $A_{j\alpha}({\bf r},z)$, which is dual to a conserved $U(1)$ current, exhibits asymptotic behavior given by $A_{\alpha} =a_{\alpha} +b_{\alpha} z + \mathcal{O}(z^2)$.
Here, $\phi$ serves as a source term and is typically set to zero in the context of the spontaneously broken symmetry phase. On the other hand, $\psi$ corresponds to the vacuum expectation value $\langle O\rangle$ of the dual scalar operator, which exhibits a non-zero value in the broken phase. The quantities $a_{x,y}$ and $b_{x,y}$ represent the superfluid velocity and the associated conjugate current, respectively. Additionally, $a_{t}$ is identified as the chemical potential $\mu$, while $b_{t}$ denotes the charge density $\rho$ of the field theory.

The charged black hole background metric employed in the Eddington coordinates takes the form of $ds^2=\frac{1}{z^2}[-f(z)dt^2-2dtdz+dx^2+dy^2]$, where $f(z)=1+Q^2(z)^4-(1+Q^2)(z)^3$, and the Hawking temperature is given by $T=(3-Q^2)/4\pi$. 
As mentioned in the main text, for a standard Schwarzschild black hole background, the conformal invariance of the superfluid implies that its thermodynamics is governed solely by the dimensionless parameter $\mu/T$, where $\mu$ represents the chemical potential and T the temperature. When the chemical potential exceeds the critical value of $4.07$, the system undergoes a transition to the broken phase as the temperature is lowered. Consequently, by fixing the characteristic chemical potential at this critical value, and varying $Q$ from $0$ to $\sqrt{3}$, we can freely control the temperature from the critical temperature down to absolute zero.

When we disregard backreaction effects and consider the complete non-equilibrium, dynamic evolution system at the boundary, we can readily trace the evolution by numerically solving the remaining equations of motion within the bulk system. Within the bulk, the equilibrium geometry is described, and the equations of motion (EOM) governing the bulk gauge and scalar fields can be expressed as follows:
\begin{align}
d_\beta F^{\alpha\beta} = J^{\alpha} = iq(\Psi^*D^{\alpha}\Psi-\Psi D^{\alpha}\Psi^*) \\
(-D^2+m^2)\Psi=0
\end{align}
By imposing the boundary conditions, these equations can be solved numerically. In particular, within a square periodic boundary, we have established the circular potential well constraints by setting the chemical potential $\mu=0$ outside the cycle with radius of 18. we employed high-order Runge-Kutta methods  with time step $\Delta t = 0.01$. Additionally, the Chebyshev method was applied in the z-direction, while the Fourier method was utilized in the x and y directions when addressing the boundary conditions with the number of grid points taken as $20 \times 300 \times 300 (nz\times nx\times ny)$ and the boundary size taken as $50\times50 (Rx\times Ry)$.

\section{2. GROSS-PITAEVSKII EQUATION}

 In the main text, we conduct simulations using the coupled Gross-Pitaevskii equations, here we introduce the Gross-Pitaevskii equation setup in detail. The Gross-Pitaevskii equation we used in main text written as
\begin{equation}
        i\hbar\frac{\partial}{\partial t}\Psi = (-\frac{\hbar^2}{2m}\nabla^2+V_{trap}+g|\Psi|^2-\mu)\Psi
\end{equation}

The order parameter is represented as $\Psi_j = \sqrt{n_j} \exp(i\phi_j)$, corresponding to the complex scalar field or order parameter as used in the holographic superfluid context. 
The interaction parameter $g$ signifies the atom interaction and is defined as $g = 2\pi\hbar^2a/m$, with $m$ the mass and $a$ the s-wave scattering length. 

Similar to holographic simulations, we have employed Fourier methods to create two-dimensional periodic boundary conditions, with the number of grid points taken as $300 \times 300 (nx\times ny)$ and the boundary size taken as $50\times50 (Rx\times Ry)$. We consider the general power-law trapping potential $V_{trap}({\bf r})=\frac{1}{2}mR^2_0(\frac{|{\bf r}|}{R_0})^\gamma$ . Then 
this potential field can be regarded as a circular infinite-depth potential well with $\gamma=100$. By using the high order Rongokuta method, we can simulate the relaxation motion of vortices in potential well.

\section{3. Finite temperature vortex relaxation dynamics}

\begin{figure}[t]
    \centering
    \includegraphics[width=11cm,trim=0 0 0 0]{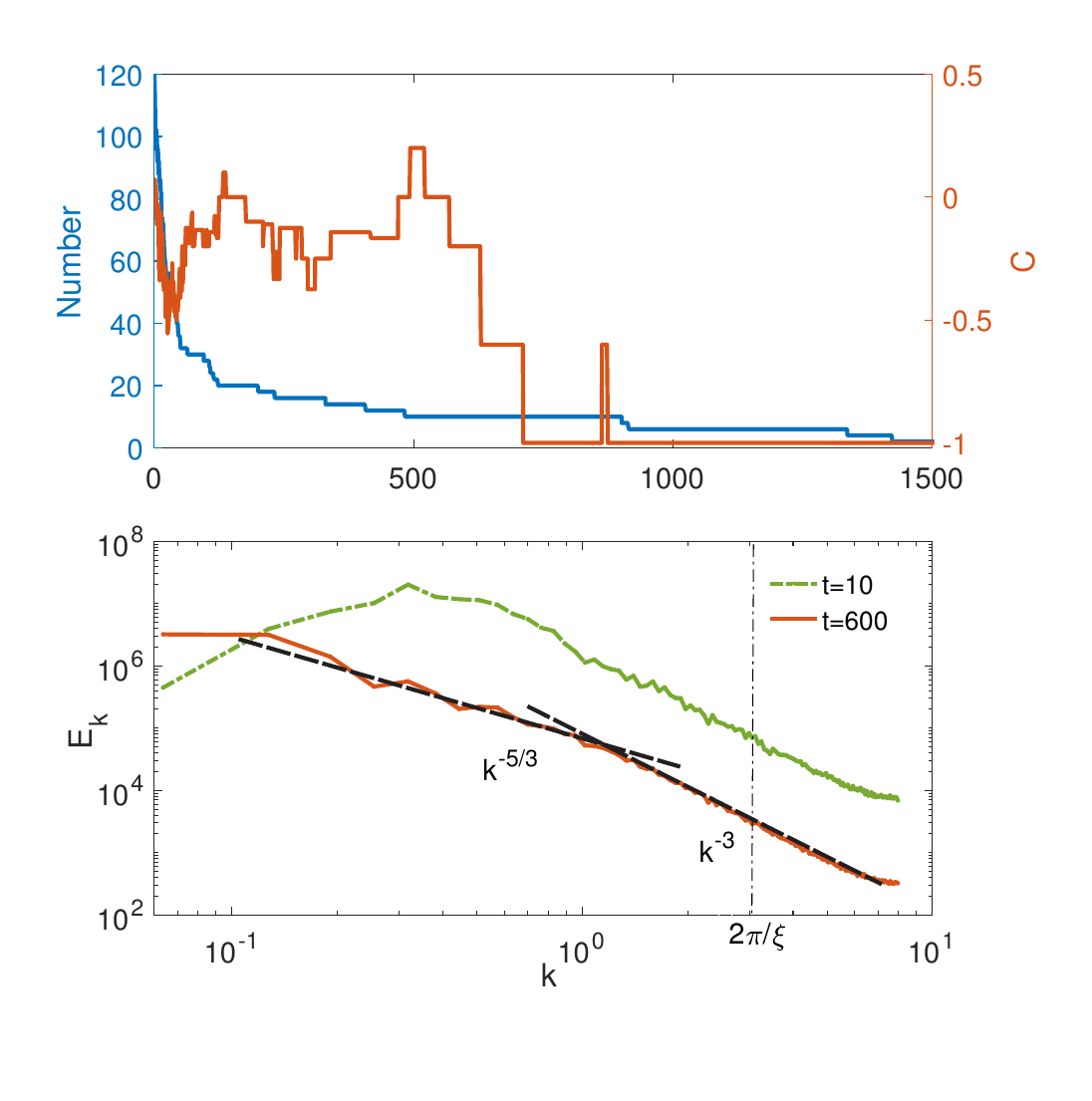}
    \caption{(a): The time evolution of vortex number and vortex cluster correlation function in the simulation of holographic model with $T=0.67T_c$, where the blue line represents the vortex number and the red line represents the correlation function.
    (b): Energy spectra for the corresponding cases at $t=10$ and $t=600$.}
    \label{figures1}
\end{figure}

In the main text, we observed the formation of stable vortex clusters in the near-zero temperature holographic superfluid, attributed to the unique dynamics of the vortices. Here, we explore the case when vortices exhibit intense motion at a finite and significantly high temperature. Considering a situation where $Q^2=1$, so the temperature is $0.67T_c$, as shown in Figure. \ref{figures1}(a). the vortices annihilate rapidly, and the correlation function $C$ rapidly decreases to $-1$ after a period of relative stability, meaning that no vortex clusters are formed.  Furthermore, as can be seen from the spectrum evolution of Figure. \ref{figures1} (b), the energy declines even at ranges greater than the healing length of the vortex, and no inverse energy cascade is observed.

\end{document}